\documentclass{aip-cp}

\usepackage[numbers]{natbib}
\usepackage{rotating}
\usepackage{graphicx}

\begin{document}

\title{Pigmy resonances, transfer, and separable potentials}
\author[aff1,aft2]{C.A. Bertulani\corref{cor1}}
\author[aft3]{A. S. Kadyrov}
\eaddress{a.kadyrov@curtin.edu.au}
\author[aff4]{A. Kruppa}
\eaddress{atk@atomki.mta.hu}
\author{T. V. Nhan Hao$^{1,}$}
\eaddress{hao.tran@tamuc.edu}
\author[aft2]{A. M. Mukhamedzhanov}
\eaddress{akram@comp.tamu.edu}
\author{Shubhchintak$^{1,}$}
\eaddress{shub.shubhchintak@tamuc.edu}
\affil[aff1]{Department of Physics and Astronomy, Texas A\&M University-Commerce, Commerce, TX 75429}
\affil[aft2]{Department of Physics, Texas A\&M University, College Station, TX, USA}
\affil[aft3]{Department of Physics and Astronomy, Curtin University, GPO Box U1987, Perth 6845,
Australia}
\affil[aff4]{Institute for Nuclear Research, Hungarian Academy of Sciences, Debrecen, PO Box 51,
H-4001, Hungary}
\corresp[cor1]{Corresponding author: carlos.bertulani@tamuc.edu}

\maketitle

\begin{abstract}
In this contribution we make a short review of recent progress on topics of current interest in nuclear physics and nuclear astrophysics. In particular, we discuss a re-analysis of the extraction of the dipole response of the pigmy resonance in $^{68}$Ni based on a continuum discretized coupled-channels calculation in relativistic Coulomb excitation experiments. We also discuss the forthcoming progresses made by our group on the Alt-Sandhas-Grassberber approach to (d,p) reactions and  future expectations. The role of separable potentials in solving such equations with a test case based on applications of such potentials to phase-shift analysis is also presented.
\end{abstract}

\section{PIGMY RESONANCES}
In the 1980's and 1990's, experiments carried out in radioactive beam facilities  lead to the belief that the observed peak in the response function due to the direct breakup of light and loosely-bound projectiles, such as $^{11}$Be and $^{11}$Li was  indicative of the existence of a collective pigmy dipole resonance (PDR) in nuclei, despite claims that such response could be well explained in a two-body breakup model  \cite{BM90}.  Heavier neutron-rich nuclei were later found to display a fragmentation of the nuclear response \cite{Lei01} in relativistic Coulomb excitation experiments \cite{Aum05, AN13}). Such experiments confirmed the existence of collective excitations at low energies in nuclei,  identified via gamma and neutron emission \cite{Lei01}. 

Fancy density functional models are now being used to describe collective states in nuclei such as the time-dependent superfluid local density approximation \cite{Bul13,Ste15,Bul16} and   pigmy resonances are often being studied using mean field theories \cite{Suz90,Vret01,Ber07,Paa07,Kre09,Pon14,Pap14}.  Slight modifications of the linear response theory allows the prediction of  a considerable concentration of the excitation strength in neutron-rich nuclei at low energies \cite{BF90,Ter91}.  The theoretical prediction of the fraction of the sum rule exhausted by pigmy resonances depends on the use of non-relativistic or relativistic mean field approaches, pairing properties, and other physical phenomena  \cite{Paa07,Kre09,Pon14,Pap14,BF90,Ter91,Col13}. The E1 strength function, is defined as  $S(E)=\sum_\nu |\left< \nu||{\cal O}_L || 0\right>|^2 \delta(E-E_\nu)$, in terms of states $\nu$, where ${\cal O}_L$ is an electromagnetic operator ($r Y_{1M}$ for dipole excitations).  The energy centroid  of the pigmy E1 strength lies probably in the range of  $7-12$ MeV for medium mass nuclei and the fraction of the sum rule exhausted can be of the order of 10\% of the total strength \cite{AGB82}.   

\begin{figure}
\includegraphics[scale=0.4]{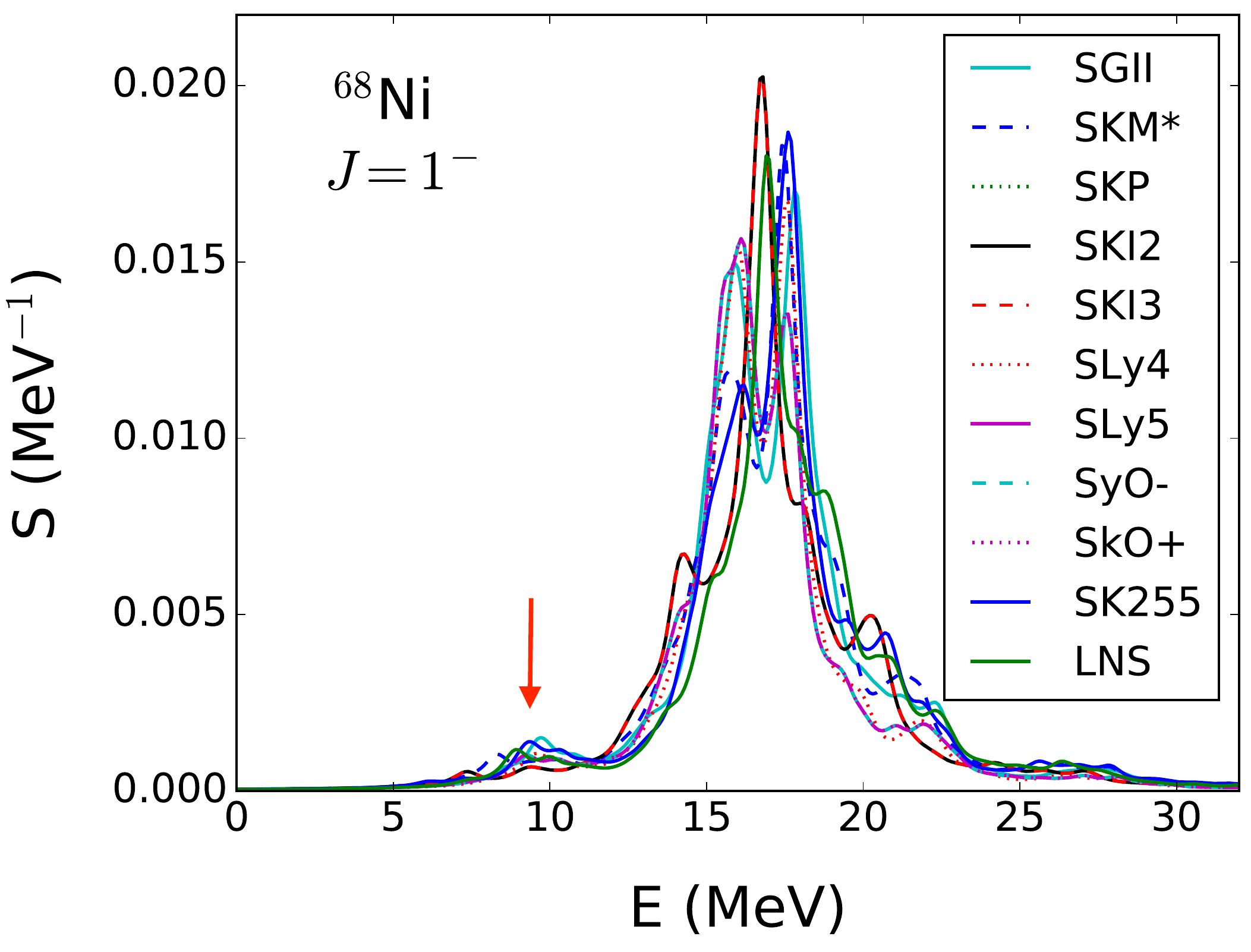}\label{figrpa}
\caption{Strength function for the E1 RPA response in $^{68}$Ni calculated with formalism described in Ref. \cite{Col13}. The calculation is performed for several Skyrme interactions. The arrow shows the prediction for location of the pygmy dipole resonance.}
\end{figure}

After a re-analysis of recent experiments, it is now well known that Coulomb excitation in heavy ion collisions at small impact parameters leads to a strong coupling of  pigmy and giant resonances and the modification of transition probabilities and cross sections  \cite{Nat16}. These effects have been observed in past experiments with regard to the coupling of giant and double giant dipole resonances  (DGDR) \cite{BB88,ABE98,BP99}. They are due to the dynamical coupling between the usual giant resonances and the DGDR  \cite{Bert96,Ber05,ABS95,PB98}.  
We have considered the excitation of  $^{68}$Ni on $^{197}$Au and $^{208}$Pb targets at  600 and 503 MeV/nucleon \cite{Nat16}, experimentally investigated in Refs. \cite{Wie09,Ros13}. The PDR in $^{68}$Ni was found at $E_{PDR} \simeq 11$ MeV with a width of $\Gamma_{PDR} \simeq 1$ MeV, exhausting approximately 5\% of the Thomas-Reiche-Kuhn (TRK) energy-weighted sum rule. But in a second experiment,  the centroid energy was observed at 9.55 MeV, exhausting  2.8\% of the TRK sum rule, and a width of 0.5 MeV. In Figure \ref{figrpa} we plot RPA calculations of the E1 response in $^{68}$Ni using ${\cal O}_{L=1} = j_1(qr)$, with $q=0.1$ fm$^{-1}$ taken as representative of the momentum transfer. We use the method descried in  Ref. \cite{Col13}. In this case, the strength function has dimensions of MeV$^{-1}$ and in the long-wavelength approximation, $qr\ll 1$, it is proportional to the usual response for electric dipole operator $rY_{1M}$. The calculation is performed for several Skyrme interactions. The arrow points to the predicted location of the PDR. Other results found in the literature display a larger response in the PDR energy region, obtained after modifications in the model space and interactions are done to accommodate the neutron-rich property of the nucleus\cite{BF90,Ter91,Paa07,Kre09,Pon14,Pap14,BF90,Ter91}.

\begin{figure}[t]
\includegraphics[scale=0.38]{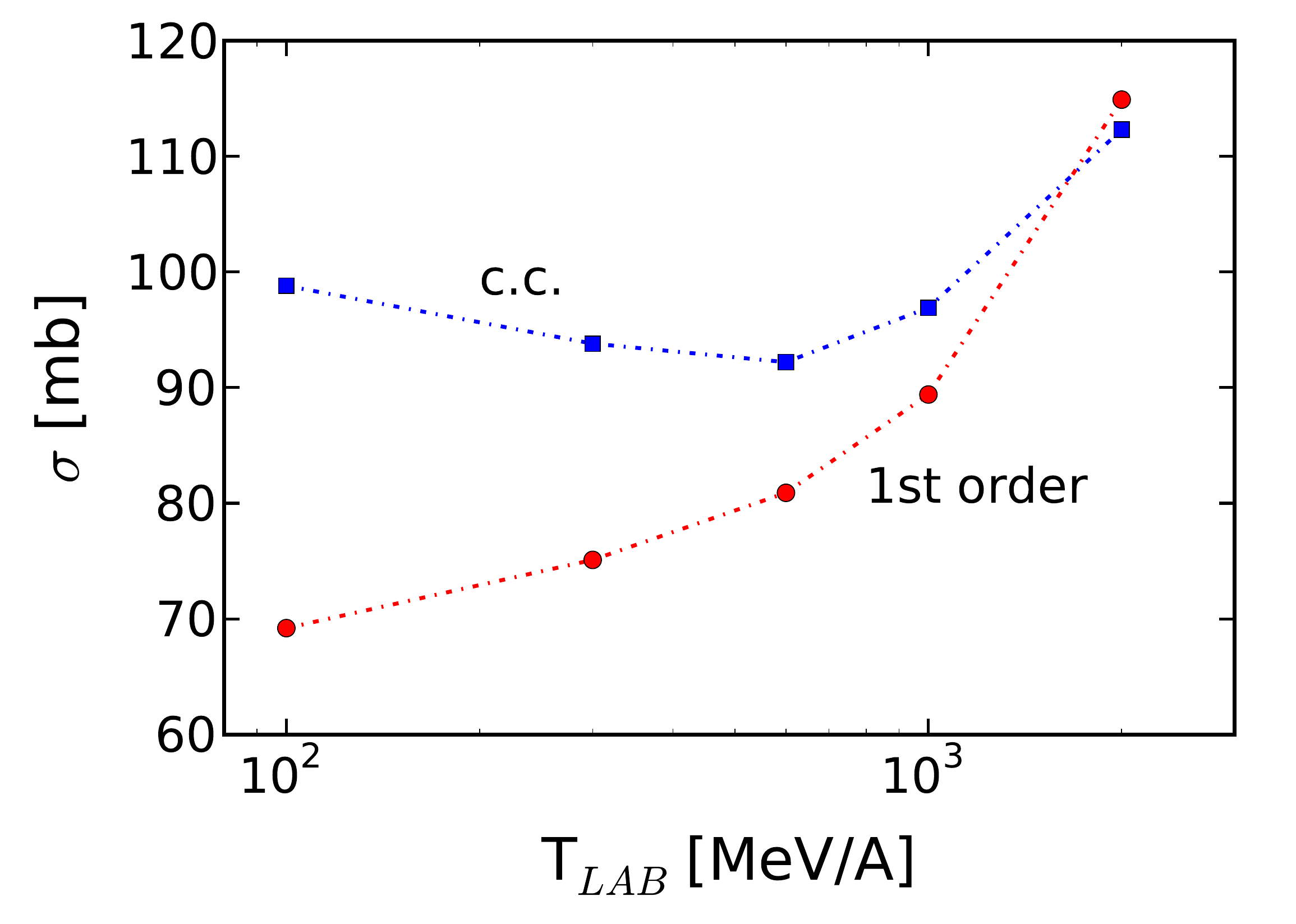}\label{CS}
\caption{Relativistic Coulomb excitation cross sections of the PDR as a function of the bombarding energy for   $^{68}$Ni projectiles incident on $^{197}$Au targets. The filled circles represent the calculations using first-order perturbation theory, while the filled squares are the results of coupled-channel calculations.}r
\end{figure}

In Ref. \cite{Nat16} one assumed Lorentzian forms for the response functions  $dB_{EL}/dE$ adjusted to a predertmined fraction of the sum-rule. This form was then used into the discretization of the response function into energy bins, yielding the the reduced matrix elements $\left|\left< \alpha || {\cal M}_{EL} || \alpha' \right>\right|^2 \propto \Delta E_x (dB_{EL}/dE)|_{E=E_x}$, where $E_x=E_{\alpha '} - E_\alpha$. Angular momentum coupling was properly taken into account (see, e.g., Ref. \cite{BP99}). More details of the calculation are found in Ref. \cite{Nat16}. In Figure \ref{CS} the results are shown for relativistic Coulomb excitation cross sections of the PDR  as a function of the bombarding energy of  $^{68}$Ni projectiles incident on $^{197}$Au targets. The filled circles represent the calculations using first-order perturbation theory, while the filled squares are the results of coupled-channel calculations following the procedure described above. At 600 MeV/nucleon the cross section for the excitation of the PDR changes from 80.9 mb using first-order perturbation theory to 92.2 mb if one uses the coupled-channels method. A re-analysis of the PDR excitation using the coupled channels method, implies an appreciable change of 14\% of the extracted PDR strength.  Its reduction by roughly this value would be needed to reproduce the experimental data.  

Calculations for  $^{68}{\rm Ni}+^{208}{\rm Pb}$ at 503 MeV/nucleon were also carried out, corresponding to the experiment of Ref. \cite{Ros13}. The Coulomb excitation cross section for the PDR in $^{68}$Ni to first-order is  found to be 58.3 mb, but when the influence of giant resonances in higher-order couplings are included, the cross section increases to 71.2 mb, a 18.1\% correction. The dipole polarizability is defined by
\begin{equation}
\alpha_D = {\hbar c\over 2 \pi^2} \int dE {\sigma(E) \over E^2},
\end{equation}
where $\sigma(E)$ is the photo-absorption cross section. The value of $\alpha_D$ extracted from the experiment in Ref. \cite{Ros13} is $3.40$ fm$^3$. But, to reproduce the experimental cross section with higher-order calculations we need only $\alpha_D = 3.16$ fm$^3$, a small but non-negligible correction. Assuming a linear relationship between the dipole polarizability and the neutron skin \cite{Pie11},  one obtains a reduction of the neutron skin from 0.17 fm, as reported in Ref. \cite{Ros13}, to 0.16 fm. This correction lies within the experimental error of 7\% for $\alpha_D$ and 0.02 fm for the neutron skin \cite{Ros13}. But such calculations highlight the fact that coupling to giant resonances must be considered in the future experimental analysis with increased data precision. This is even more important for experiments at lower bombarding energies.

\section{TRANSFER REACTIONS}
Theoretically, deuteron induced-reactions (d, p) is the simplest
transfer reactions which still carry many of the basic features of rearrangement reactions induced by heavier nuclei. They are also attractive from the experimental perspective since the deuterated targets are readily available. In inverse kinematic, (d, p) and (d, n) reactions provide an useful tool for extracting nuclear information for neutron capture. Such neutron cross sections are clearly identified as one of the crucial inputs needed for nuclear astrophysics and applied physics. Reaction theory is the key to connecting (d, p) cross sections to neutron capture cross sections. In the 1960s and 1970s (d, p) reactions were studied in great detail. Later (d, p) reactions have been recycled on the scientific stage, benefiting from huge progress in experimental techniques, especially in the development of radioactive beam facilities. However, the standard analysis of (d, p) data for more than 50 years still relies on the distorted wave Born approximation (DWBA). The theoretical description of reactions in general, and of the theory for (d, p) reactions in particular, needs to advance into the new century.

In principle, scattering of deuteron off
a nuclear target (an inert core $A$) can  be exactly described in the framework of the 3-body Faddeev integral equations written in the AGS form \cite{AGS67}
\begin{equation}\label{AGS}
  U_{\beta \alpha}= \bar{\delta}_{\beta \alpha}G^{-1}_0+\sum^{3}_{\gamma=1}\bar{\delta}_{\beta\gamma}T_{\gamma}G_0U_{\beta\gamma}
\end{equation}
where $U_{\beta \alpha}$ is the transition operator whose on-shell matrix elements $\langle \Phi_{\beta}| U_{\beta \alpha}|\Phi_{\alpha}\rangle$ are the scattering amplitudes of the transition from the initial channel $\alpha +(\beta \gamma)$ to the final channel $\beta+(\alpha \gamma)$, $\bar{\delta}_{\beta \alpha}=1-\delta_{\beta \alpha}$ is the anti-Kronecker symbol, $G_0=(E+i0-H_0)^{-1}$ is the free resolvent, $E$ is the available three-particle energy in the center of mass system, $H_0$ is the free Hamiltonian, $T_{\gamma}=V_{\gamma}+V_{\gamma}G_{\gamma}V_{\gamma}$ is the transition operators of the two-particles systems, and $V_{\gamma}$ the potential for the pair $\gamma$ in odd-man-out notation. The channel states $|\Phi_{\gamma}\rangle$ are the eigenstates of the corresponding channel Hamiltonian $H_{\gamma}=H_0+V_{\gamma}$.

\begin{figure}[tbp]
\includegraphics[width=10.0cm]{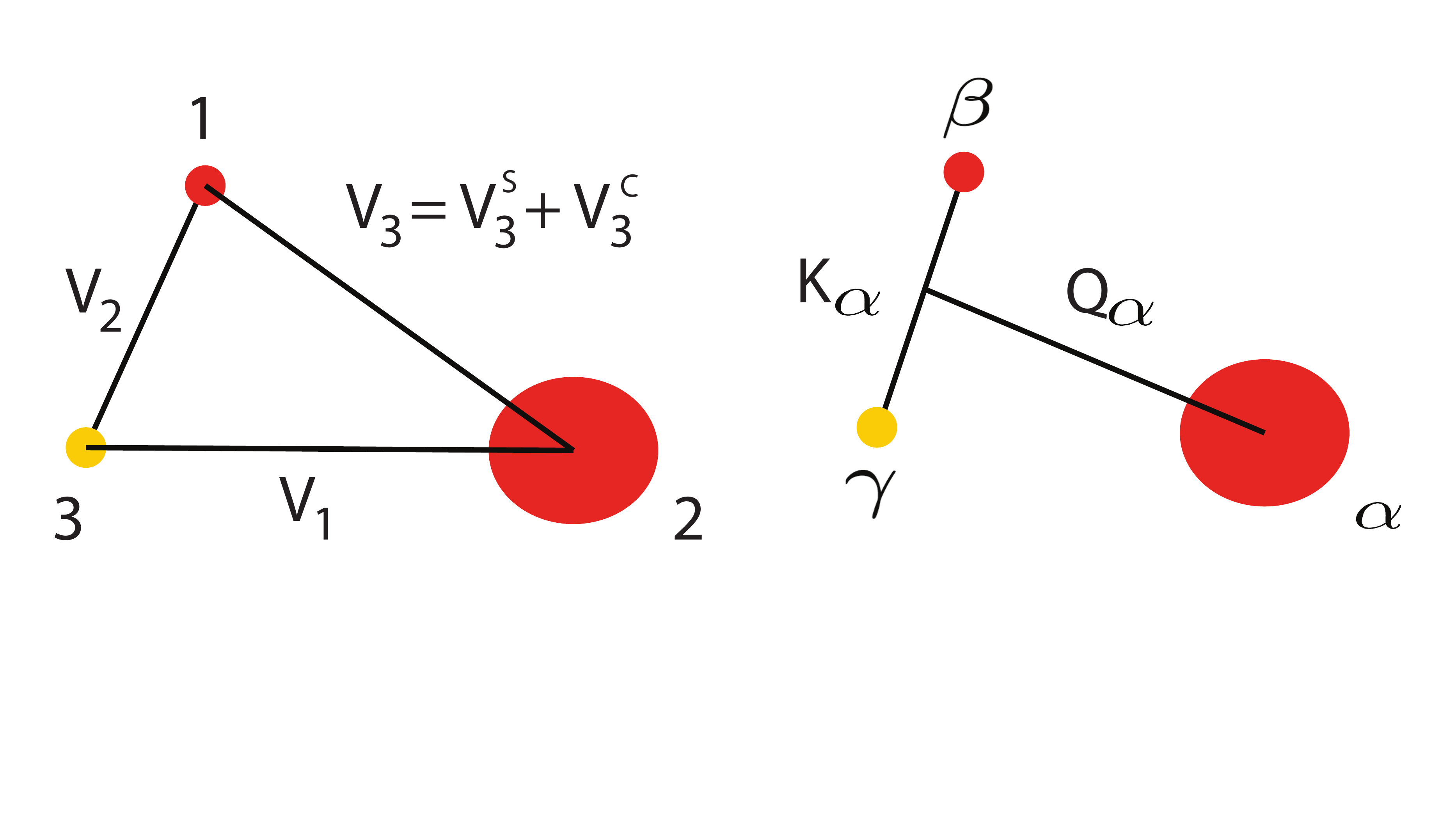}
\caption{{(Left) Sketch of the three-body system and the labels for them and for their mutual interactions. $V_3$ is the sum of nuclear [$V_3^{S}$ (short range)] and Coulomb ($V_3^{C}$) interaction between proton (1) and nucleus (2), whereas $V_1$ and $V_2$ are short range nuclear interactions of nucleus (2) and the neutron (3) and between the proton (1) and neutron (3), respectively.  (Right) Labels for kinetic energy operators for the same system.
}}
\label{fig_hamiltonian}
\end{figure}

If the internal structure of the target can no longer be neglected, the generalization of the AGS equations
is achieved by taking into account the excitation of the
target \cite{alt2007,MES12}. The Hamiltonian (see Fig. \ref{fig_hamiltonian}) is given as
\begin{eqnarray} \label{3Htotal}
{\it H}={\it H}_{\mathrm{int}}+{\it H}_0+{\it V}, \ \ \ \ \  {\rm where} \ \ \
{\it H}_0={\it{\mathbf{K}}}^2_\alpha /2\mu_\alpha + {\it{\mathbf{Q}}}^2_\alpha /2 M_\alpha,
\ \ \ \  \ {\rm and}  \ \ \
{\it V}={\it V}_1+{\it V}_2+{\it V}_3.
\end{eqnarray}
${\sl H}_{\mathrm{int}} = T_{\mathrm{int}} + V_{\mathrm{int}}\,$ is the internal Hamiltonian of nucleus $2$; $\,T_{\mathrm{int}}$ and $V_{\mathrm{int}}$ are the internal kinetic energy operator and  the internal potential of nucleus $2$. ${\it H}_0$ is the Hamiltonian of the relative motion of the non-interacting particles 1, 3 and the center of mass of particle 2. That is, ${\it {\mathbf{K}}}_\alpha$ is the momentum operator for the relative motion of particles $\beta$ and $\gamma$ and $\mu_\alpha=m_\beta m_\gamma/m_{\beta \gamma}$ the corresponding reduced mass, $m_{\beta\gamma}= m_{\beta} + m_{\gamma}$; $\,\,{\it {\mathbf{Q}}}_\alpha$ is the relative momentum operator for the motion of particle $\alpha$ and the center of mass of $(\beta,\gamma)$ with $M_\alpha=m_\alpha\,m_{\beta\gamma}/(m_\alpha+m_\beta+m_\gamma)$. The potentials ${\it V}_1$ and ${\it V}_3$ describe the interaction of the nucleons 1 and 3, respectively, with each of the constituents of nucleus 2, and ${\it V}_2$ is the internucleon potential.  The potential ${\it V}_3 = {\it V}_3^{S}  + {\it V}_3^{C}$, where ${\it V}_3^{S}$  and ${\it V}_3^{C}$  are the short-range and the Coulomb part of the proton target interaction, respectively.

Consider the case that the target can exist in several internal states $\rho$ ($\rho=1,2\dots ,N$), assumed to be orthogonal, with wave functions $\left|\varphi^\rho\right>$ and energies $\epsilon^\rho \geq 0$, that is,
\begin{equation}
{\it H}_{\mathrm{int}}\left|\varphi^\rho\right>=\epsilon^\rho \left|\varphi^\rho\right>. \label{targetexci}
\end{equation}
The notation is such that $\rho =1$ corresponds to the ground state with $\epsilon^1=0$. The index $\rho$ is supposed to contain the complete specification of the internal state, in particular also its spin, isospin etc. To reduce the (A + 2)-particle problem to the much
simpler three-body problem all the operators acting in the
(A + 2)-particle space were projected onto the three-particle
space. In this way they become $N \times N$ matrix operators
\begin{equation}
{\b H} =[H^{\rho\sigma}]=[\left<\varphi^\rho |{\it H}| \varphi^\sigma \right>],
\end{equation}
\begin{equation}
{\b H}_0=[H_0^{\rho\sigma}]=[\left<\varphi^\rho |{\it H}_0|
\varphi^\sigma \right>]=
[\delta_{\rho\sigma}({\it {\mathbf{K}}}^2_\alpha /2\mu_\alpha +{\it {\mathbf{Q}}}^2_\alpha /2
M_\alpha)],
\end{equation}
\begin{equation}
{\b V}_\alpha=[V_\alpha^{\rho\sigma}]=[\left<\varphi^\rho |{\it V}_\alpha|
\varphi^\sigma \right>],
\end{equation}
and $[\left<\varphi^\rho |{\it {\mathbf{Q}}}_\alpha| \varphi^\sigma \right>]=[\delta_{\rho\sigma}\,{\mathbf{Q}}_\alpha]$ and $[\left<\varphi^\rho |{\it {\mathbf{K}}}_\alpha| \varphi^\sigma \right>]=[\delta_{\rho\sigma}\,{\mathbf{K}}_\alpha]$. The resolvent matrices corresponding to the restricted full and free Hamiltonian matrices are
\begin{eqnarray}
{\mathcal{G}}(z)=(z-{\b H})^{-1},
\ \ \ \ \ \ {\rm and} \ \ \ \ \ {\mathcal{G}}_0(z)=(z-{\b H}_0)^{-1}.
\end{eqnarray}
All the operators acting in the $A+2$ space and projected onto the A-body space are underlined, while their matrix elements, which have upper indices characterizing the excited states of nucleus $A$, are not. Now we can introduce the modified transition operators satisfying the AGS equations (the Coulomb interaction between particles $1$ and $2$ is, for the moment, disregarded). As shown in Ref. \cite{MES12}, the modified transition operators $\mathcal{U}_{\beta\alpha}(z)$ also satisfy the AGS equations

\begin{equation}
\underline{\mathcal{U}}_{\beta\alpha}(z)={\bar {\delta}}_{\beta\alpha}\underline{\mathcal{G}}^{-1}_0(z) +\sum_\gamma \bar {\delta}_{\gamma\alpha} \underline{\mathcal{U}}_{\beta\gamma}(z)\underline{\mathcal{G}}_0(z)\underline{t}_\gamma(z),
\label{AGSmod1}
\end{equation}
where $\underline{\mathcal{U}}$ is the three-body modified transition operators, $\underline{\mathcal{G}}_0(z)$ is the three-body modified free Green's function, $\underline{t}_{\gamma}(z)$ is the two-body T operators in the three-body space for subsystems (p+A) or (n+A) via the Lippmann-Schwinger equation

\begin{equation}
{\b t}_\alpha(z)={\b V}_\alpha+{\b V}_\alpha \underline{\mathcal{G}}_0(z){\b t}_\alpha(z).
\label{LSM}
\end{equation}

\begin{figure}[t]
\includegraphics[width=18.0cm]{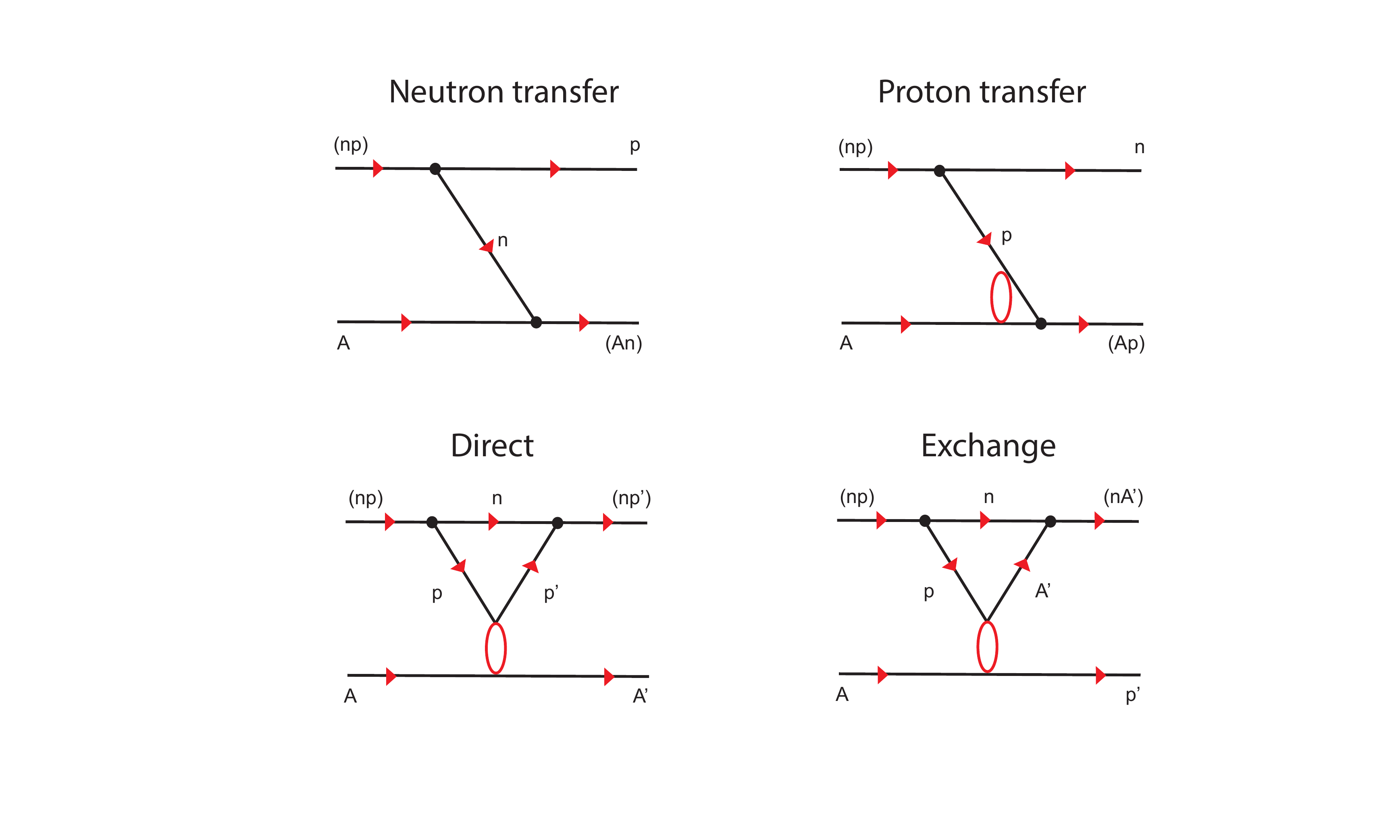}
\caption{Diagrammatical representation of various channels in (d,p) reactions in the framework of the 3-body Faddeev/AGS equations. The bubbles represent
the non-perturbative off-shell Coulomb scattering amplitudes of proton and target.} \label{potentials}
\end{figure}

The generalized AGS equations obtained above couple all rearrangements in the inelastic and elastic amplitudes (see Fig. \ref{potentials}). The transition amplitudes for all interesting three-body processes are obtained, whether the target nucleus is in its ground or in some excited state before and/or after the collision.

The AGS equations are usually solved in momentum-space partial-wave basis where they become a system of integral equations with two continuous momenta variables. It is well known that the AGS equations could only be solved for short-range potentials $V_{\gamma}$. However, at low-energy the AGS equations can reproduce the data accurately if the long-range Coulomb interaction $V^{C}_3$ is taken into account. One way of dealing with the problem is to use the method of screening and renormalization which enables to calculate the Coulomb-distorted short-range part of the transition amplitude by solving the AGS equations with nuclear plus screened Coulomb potential. The amplitude calculated with the screened potential will be renormalized by dividing with the Coulomb renormalization factor. The physical amplitude will be obtained by repeating the calculation for successively increasing values of the radial distance $R$. Within this approach, the current implementation of the AGS equations has very successful in describing the light nuclear systems \cite{alt2002,deltuva2009,deltuva2009b}. However, the technical difficulties arise in the renormalization procedure as the charge of the target increases making this approach unreliable for targets with charge $Z\geq20$. Again, the treatment of the long-range character of the Coulomb interaction in low-energy nuclear reactions remains an interesting and challenging task. An alternative approach based on employing the non-screened Coulomb potential was proposed in Ref. \cite{MES12}. Instead of the usual plane-wave basis, the AGS equations are cast in a momentum-space Coulomb distorted partial-wave representation. Applying two potential formulas the AGS equations are converted to the form in which the matrix elements are sandwiched by the Coulomb distorted waves in the initial and final states. Then the generalized AGS integral equations were formulated for the first time with explicit inclusion of the Coulomb interaction and target excitations (as mentioned above, the original AGS equations were formulated for three structureless particles). The equations provide the most advanced and complete description of the deuteron stripping. Despite the core formalism having been developed, in order to make it practical an important additional analytical and computational work is required and is currently under investigation.

\section{RADIATIVE CAPTURE REACTIONS}

Eq. (\ref{AGSmod1}), which is a coupled equation and is the solution of AGS equations is difficult to solve. However, it could be simplified if the transition operator $\underline{t}_{\gamma}(z)$ are represented in separable form. This can be done by writing two particle short range interactions in terms of (quasi-) separable potentials. In fact, short range nuclear interactions by nature are non-local and can be easily handled with separable potentials.  For the nucleon-nucleus subsystem $\alpha(=1,3)$, i.e., for the neutron-target and proton-neutron system, one can write the two body potential in Eq. (6) in separable form as,
\begin{eqnarray}
V_{\alpha}^{\rho \sigma} = \sum_{t_{\alpha} t'_{\alpha}}^{A_{\alpha}}|\chi_{\alpha m}^{\rho}\rangle \lambda_{\alpha;mn}^{\rho \sigma} \langle \chi_{\alpha n}^{\sigma}|, \label{s1}
\end{eqnarray}
where, $m(n)$ collectively represents the quantum numbers of the two-body state before (after) the interaction and is called the reaction channel.  $t_{\alpha}$ ($t'_{\alpha})$ are the numbers of separable expansion terms before (after) the interaction. The total number of expansion terms $A_{\alpha}$ could be greater than the number of the bound states $N_\alpha$ in the pair $\alpha$, where the terms with $t_{\alpha} > N_{\alpha}$ represent auxiliary terms which indeed are not the bound states but are added to improve accuracy.  $|\chi_{\alpha m}^{\rho}\rangle$ is the form factor vector in channel $m$ having target internal excitation $\rho$, whereas $\lambda_{\alpha;mn}^{\rho \sigma}$ is the coupling matrix which is chosen as symmetric to ensure the hermiticity of the potential (i.e. $\lambda_{\alpha;mn}^{\rho \sigma} = \lambda_{\alpha;mn}^{\sigma \rho}$). Furthermore, $\lambda_{2;mn}^{\rho \sigma} = \delta_{\rho \sigma}\lambda_{2;mn}$, as the internal state of the target does not effect the nucleon-nucleon interaction.
For further details and details on simplifications of Eq. (\ref{AGSmod1}) with  quasi-separable potentials,  Eq. (\ref{s1}), one is referred to Ref. \cite{cattapan1}. The above representation of the separable potential is bit complicated and to understand the utility of separable potentials in more details we will discuss few examples of nucleon-nucleus scattering using rank one potentials without core excitation.

\begin{figure}[t]
\centering
\includegraphics[trim=0.5mm 0.5mm 0.5mm 0.5mm,clip,width=7cm]{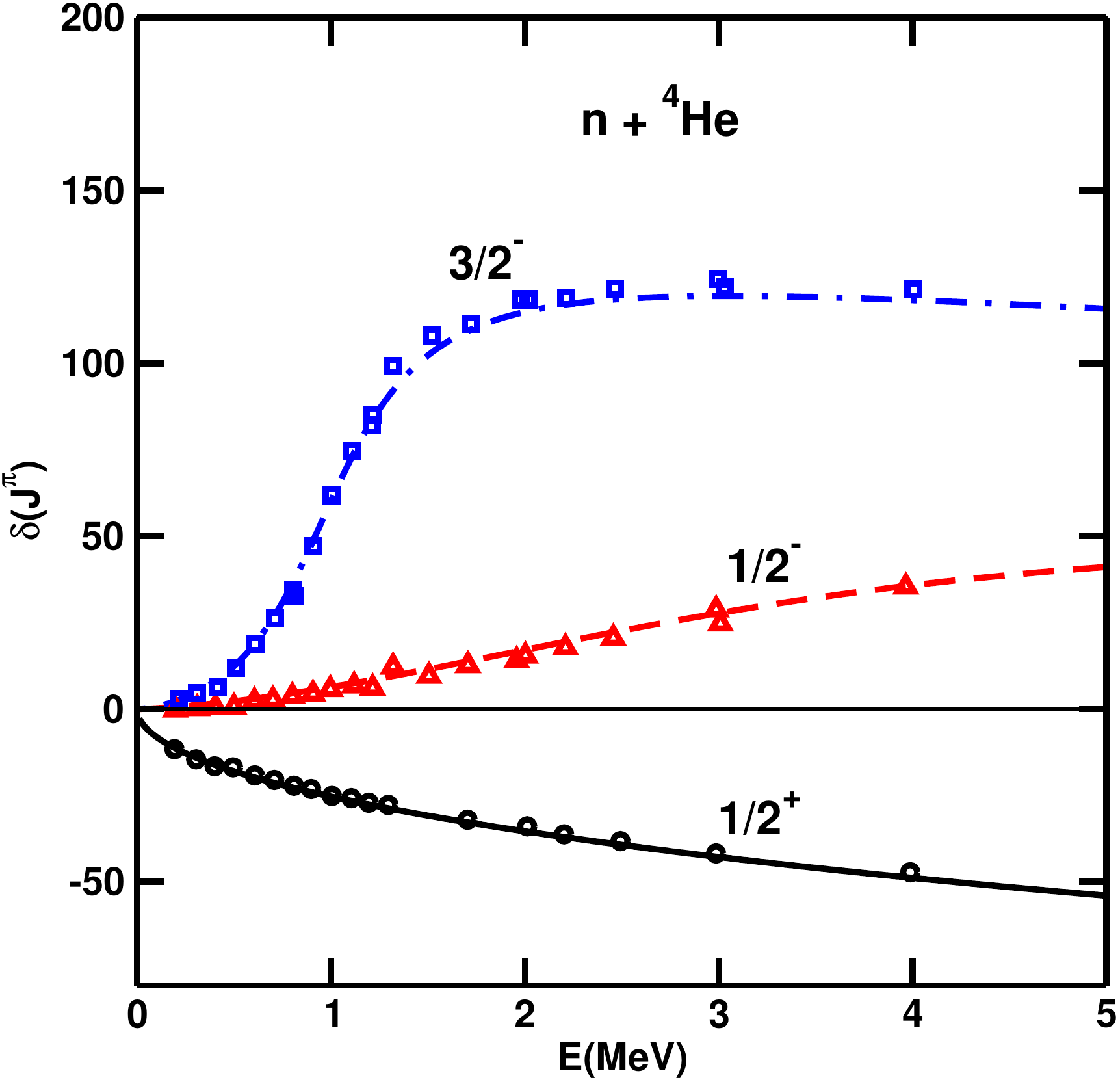}
\includegraphics[trim=0.5mm 0.5mm 0.5mm 0.5mm,clip,width=7cm]{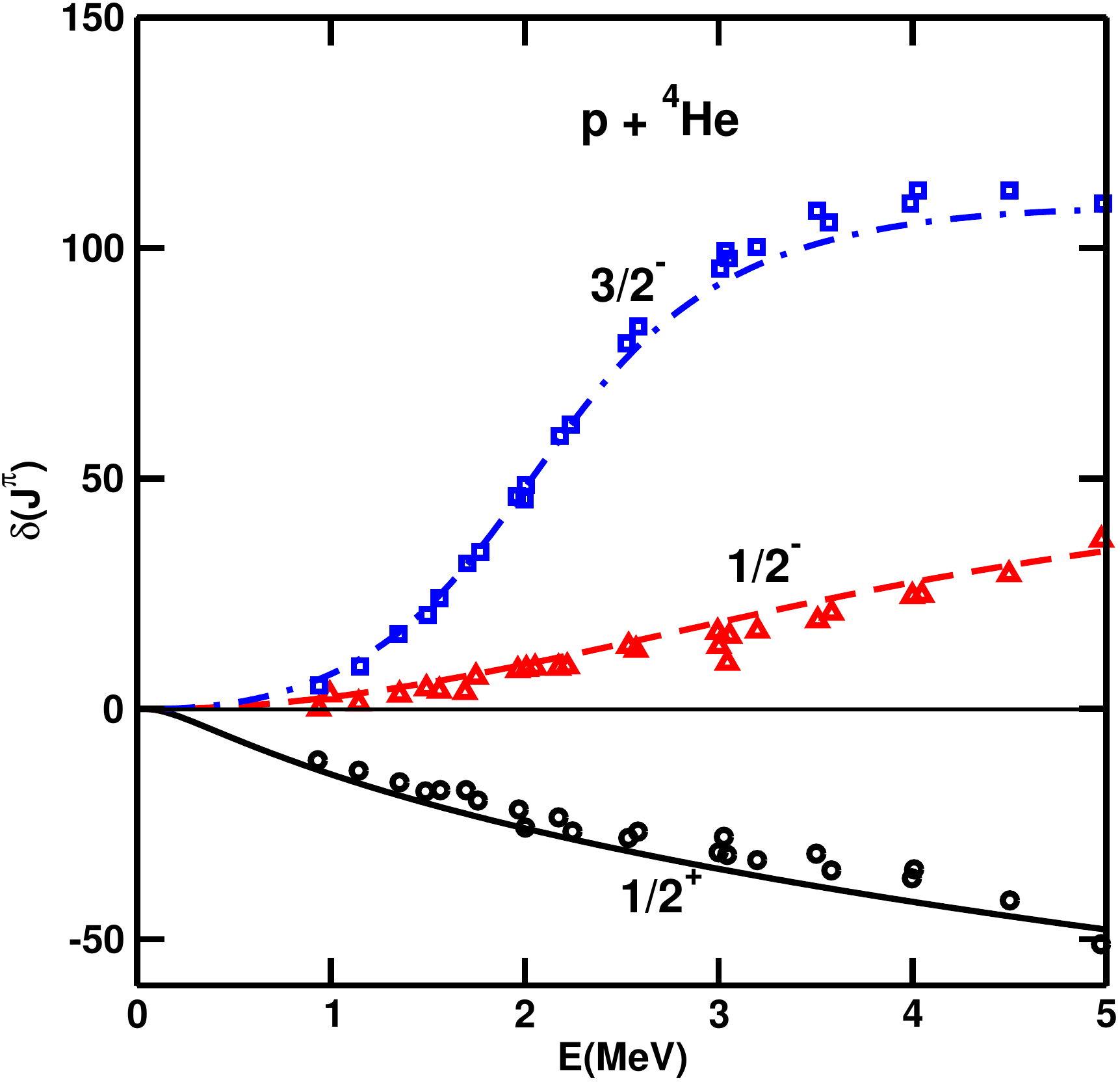}
\caption{\label{fig:1} $s$ and $p$ phase shifts for n-$^{4}$He (left panel) p-$^{4}$He and scattering . Solid, dashed and dotted-dashed lines represent the calculated $s_{1/2}$, $p_{1/2}$ and $p_{3/2}$ phase shifts, respectively, whereas filled circles, upper triangles and square boxes represent the corresponding experimental data which are taken from Refs. \cite{a1,a2,a3,a4}.}
\end{figure}

In fact, in the literature a large number of separable potentials are available which have been found to work well in different energy regions and to depend on the relative angular momentum state of the scattering particle (for eg. see Refs. \cite{cattapan1,cattapan2,miyagawa, alt_02}). We will consider the one given in Refs. \cite{cattapan1,cattapan2}, which works well for $l \leq 2$ in the low energy range (below 5-7 MeV). 

Consider the nucleon-nucleus scattering, with $I_m$ being the spin of target in channel $m$ (with the convention that $m = 1$ represents the elastic channel), where the state of the system in partial wave representation is denoted as $|m l j J M\rangle$ (note that here notation $m$ is used to represent the channel). $l$ is the relative angular momentum of the nucleon with respect to the target which couples with the nucleon spin ($s$) to yield $j$. This then couples with $I_m$ to give the total angular momentum $J$ of the system, with $M$ being its projection. Then the simplified form of Eq. (\ref{s1}), similar to the one in Ref. \cite{cattapan2}, in terms of partial wave representation is,
\begin{eqnarray}
V = \sum_{m l j,n l' j';JM}|{\bf \chi};m l j JM\rangle \lambda_{m l j,n l' j'}^{J\Pi}\langle n l' j' JM;{\bf \chi}|. \label{s2}
\end{eqnarray}
The form factor vector $|{\bf \chi};m l j JM\rangle$ in the coordinate and momentum space is represented as
\begin{eqnarray}
\langle m l j JM;{r}|{\bf \chi}; n l' j' J'M'\rangle \nonumber &=&  \delta_{m n}\delta_{l l'}\delta_{jj'}\delta_{JJ'}\delta_{MM'} i^{-l}v_{m l j}^J(r),
~~~~ \label{a4}
\\
\langle m l j JM;{k}|{\bf \chi}; n l' j' J'M'\rangle \nonumber &=&  \delta_{m n}\delta_{l l'}\delta_{jj'}\delta_{JJ'}\delta_{MM'} i^{-l}u_{m l j}^J(k). \label{a5}
\end{eqnarray}
The form factor $v_{m l j}^J(r)$ used in Refs. \cite{cattapan1,cattapan2} is taken as a  modified-Yukawa type,
\begin{eqnarray}
v_{m l j}^J(r) = \exp(-\mbox{\boldmath$\beta$}_{m lj}^J r) r^{l-1}, \label{a6}
\end{eqnarray}
and
\begin{eqnarray}
u_{m l j}^J(k) = \sqrt{\frac{2}{\pi}}\frac{1}{k}\int_0^{\infty} F_l(kr)v_{m l j}^J(r)rdr. \label{a7}
\end{eqnarray}

\begin{figure}[t]
\centering
\includegraphics[trim=0.5mm 0.5mm 0.5mm 0.5mm,clip,width=7cm]{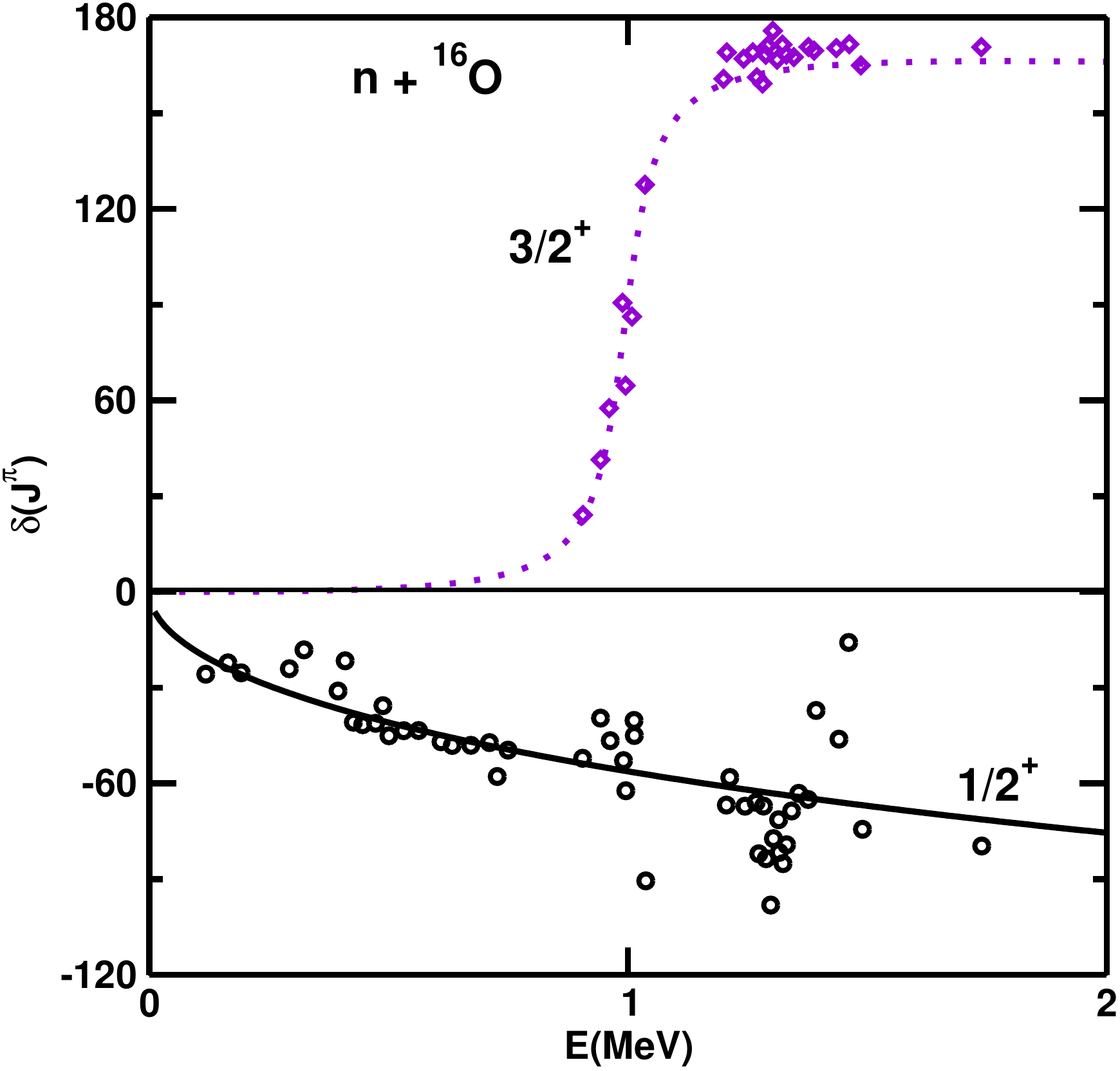}
\includegraphics[trim=0.5mm 0.5mm 0.5mm 0.5mm,clip,width=7cm]{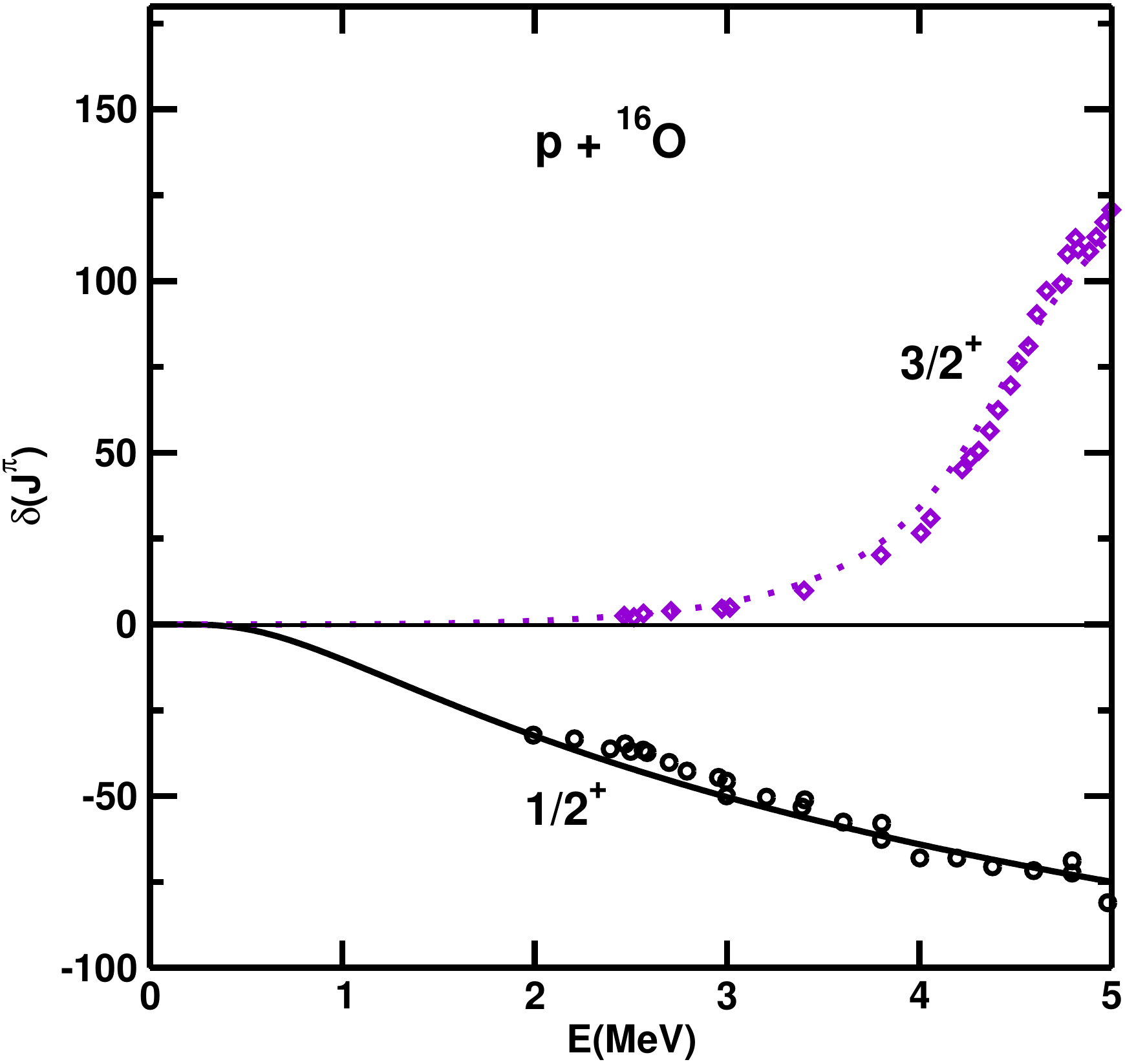}
\caption{\label{fig:2} $s$ and $d$ phase shifts for n-$^{16}$O and p-$^{16}$O scattering. Solid and dotted lines represent the calculated $s_{1/2}$ and $d_{3/2}$ phase shifts, respectively, whereas filled circles and diamond boxes represent the corresponding experimental data which are taken from Refs. \cite{a5,a6,a7,a8,a9,a10}.}
\end{figure}

Here $\mbox{\boldmath$\beta$}_{m lj}^J$ is a fitting parameter having units of inverse of length and $\mbox{\boldmath$\beta$}_{m lj}^J  = \mbox{\boldmath$\beta$}_{lj}^J$. In order to determine the parameters $\lambda$ and $\mbox{\boldmath$\beta$}$ and hence the potential, one needs to fit the experimental phase shifts which can be calculated from the elastic S-matrix.

Using the potential (\ref{s2}), the S-matrix can be written in the form:
\begin{eqnarray}
{S}_{m l j,n l' j'}^{J\Pi} &=& \delta_{m n}\delta_{ll'}\delta_{jj'} + i^{l'-l+1}\pi \mu k_{m}^{1/2}u_{m l j}^J(k_{m}){\cal T}_{m l j,n l' j'}^{J\Pi}u_{n l' j'}^J(k_{n})k_{n}^{1/2},\label{a6}
\end{eqnarray}
where, $\mu$ is the reduced mass of the two particle system. The ${\cal T}_{m l j,n l' j'}^{J\Pi}$ is a matrix related to two body T-matrix (t) and satisfying the following coupled equation,
\begin{eqnarray}
{\cal T}_{m l j,n l' j'}^{J\Pi} = \lambda_{m l j,n l' j'}^{J\Pi} - \sum_{s l'' j''} \lambda_{m l j,s l'' j''}^{J\Pi} {\cal G}_{s l'' j''}^{J\Pi}{\cal T}_{s l'' j'',n l' j'}^{J\Pi}, \label{a8}
\end{eqnarray}
with
\begin{eqnarray}
{\cal G}_{s l'' j''}^{J\Pi} &=& \langle s l'' j'' JM;{\bf \chi}|G_0|{\bf \chi};s l'' j'' JM\rangle 
=\mu\int_0^{\infty}\frac{[u_{s l''j''}(k)]^2 k^2}{k_{s}^2-k^2+i\epsilon}dk. \label{a9}
\end{eqnarray}
The calculations of the S-matrix for proton-nucleus collision is similar but it involves Coulomb modified form factors. For further details, see Refs. \cite{cattapan1,cattapan2}.

The advantage of this potential is that the fitting parameters needed for the proton-nucleus scattering are almost the same as those needed for neutron-nucleus scattering. The above procedure is given for the case when there are more than one target state involved. But for the case of light nuclei, even the single channel calculations gives good results \cite{Shub16}. This is also clear from the Figures 1 and 2, where we have calculated the phase shifts using only the ground state of the target. In Fig. 1, the left and right panels show the $s_{1/2}$ (solid), $p_{1/2}$ (dashed) and $p_{3/2}$ (dotted-dashed lines) phase shift for n-$^4$He and p-$^4$He scattering, respectively. The experimental data for left panel are from Refs. \cite{a1,a2} and for right panel are from Refs. \cite{a3,a4}, respectively. The calculations match well with the experimental data for the parameters given in Ref. \cite{cattapan1}. In, Fig. 2, we have plotted the $s_{1/2}$ (solid) and $d_{3/2}$ (dotted) phase shifts for the n-$^{16}$O and p-$^{16}$O scattering, in the left and right panels respectively. In this case, the experimental data are from Refs. \cite{a1,a5,a6,a7,a8,a9,a10}.

Note that the single channel potential can reproduce the phase shifts only in light nuclei and with the increase of the nuclear mass, the nuclear structure becomes more complicated and one needs to take into consideration the effect of other channels in order to explain the resonances which also become of non-single-particle nature. 
 
\section{CONCLUSIONS}

In this short review we have shown that the extraction of the dipole strength of pigmy resonances from relativistic Coulomb excitation experiments need a careful treatment of the dynamical coupling with giant resonance states.

We have also described a procedure to obtain phase-shifts with separable potentials which will be used for the computation of (d,p) reactions of interest to radiative neutron capture in stars. A recent work has made use of this procedure to study radiative capture  capture reactions of interest for nuclear astrophysics \cite{Shub16}. 

\section{ACKNOWLEDGMENTS}
C.A.B. acknowledges support by US National Science Foundation Grant No. 1415656 and US Department of Energy Grant No. DE-FG02-08ER41533. A.M.M. acknowledges  support by the US Department of Energy Grant No. DE- FG02-93ER40773, the US Department of Energy, National Nuclear Security Administration Grant No. DE-FG52-09NA29467, and the US National Science Foundation Grant No. PHY-1415656. 

\bibliographystyle{aipnum-cp}
\bibliography{sample}

\begin{thebibliography}{99}
\bibitem{BM90} C.A. Bertulani and M.S. Hussein, Phys. Rev. Lett. 64, 1099 (1990).
\bibitem{Lei01} A. Leistenschneider et al., Phys. Rev. Lett. 86, 5442 (2001).
\bibitem{Aum05} T. Aumann, Phys, Eur. J A 26, 441 (2005).
\bibitem{AN13} T Aumann and T Nakamura, Phys. Scr. T152, 014012 (2013).
\bibitem{Bul13} A. Bulgac, Annu. Rev. Nucl. Part. Sci. 63, 97 (2013).
\bibitem{Ste15} I. Stetcu, C.A. Bertulani, A. Bulgac, P. Magierski, K.J. Roche Phys. Rev. Lett. 114, 012701 (2015).
\bibitem{Bul16} Aurel Bulgac, Piotr Magierski, Kenneth J. Roche, Ionel Stetcu, Phys. Rev. Lett. 116, 122504 (2016).
\bibitem{Suz90} Y. Suzuki, K. Ikeda, and H. Sato, Prog. Theor. Phys. 83, 180 (1990).
\bibitem{Vret01} D. Vretenar et al., Nucl. Phys. A 692, 496 (2001).
\bibitem{Ber07} C.A. Bertulani, Phys. Rev. C 75, 024606 (2007); Nucl. Phys. A 788, 366 (2007).
\bibitem{Paa07} N. Paar, D. Vretenar, E. Khan, and G. Col\`o,  Rept. Prog. Phys. 70, 691 (2007).
\bibitem{Kre09} S. Krewald and J. Speth,  Int. J. Mod. Phys. E18, 1425 (2009).
\bibitem{Pon14} V.Yu. Ponomarev, J. Phys: Conf. Series 533, 012028 (2014).
\bibitem{Pap14} P. Papakonstantinou, H. Hergert, V.Yu Ponomarev, and R. Roth,  Phys. Rev. C 89, 034306 (2014).
\bibitem{BF90} G. Bertsch and J. Foxwell, Phys. Rev. C41 (1990) 1300. (Erratum: Phys. Rev. C42, 1159 (1990).
\bibitem{Ter91} N. Teruya, C.A. Bertulani, S. Krewald, H. Dias and M. S. Hussein, Phys. Rev. C 43, 2049 (1991).
\bibitem{Col13} G. Col\`o, L. Cao, N. V. Giai and L. Capelli, Comput. Phys. Comm. 184, 142 (2013).
\bibitem{AGB82} Y. Alhassid, M. Gai, and G.F. Bertsch, Phys. Rev. Lett. 49,1482 (1982).
\bibitem{Nat16} N.S. Brady, T. Aumann, C.A. Bertulani, and J.O. Thomas, Phys. Lett. B 757, 553 (2016).
\bibitem{BB88} C.A. Bertulani and G. Baur, Phys. Reports 163, 299 (1988).
\bibitem{ABE98} T. Aumann, P. F. Bortignon, and H. Emling, Annu. Rev. Nucl. Part. Sci. 48, 351 (1998).
\bibitem{BP99} C.A. Bertulani and V. Ponomarev, Phys. Reports 321 (1999).
\bibitem{Bert96} C.A. Bertulani, L.F. Canto, M.S. Hussein and A.F.R. de Toledo Piza, Phys. Rev. C 53, 334 (1996). 
\bibitem{Ber05} C. A. Bertulani, Phys. Rev. Lett. 94, 072701 (2005).
\bibitem{ABS95} T. Aumann, C.A. Bertulani and K. Suemmerer, Phys. Rev. C 51, 416 (1995). 
\bibitem{PB98} V.Yu. Ponomarev and C.A. Bertulani, Phys. Rev. C 57 (1998) 3476.
\bibitem{Wie09} O. Wieland et al., Phys. Rev. Lett. 102, 092502 (2009).
\bibitem{Ros13} D. M. Rossi at al., Phys. Rev. Lett. 111, 242503 (2013).
\bibitem{AGS67} E. O. Alt, P. Grassberger, and W. Sandhas, Nucl. Phys. B2, 167 (1967).
\bibitem{alt2007} E. O. Alt, L. D. Blokhintsev, A. M. Mukhamedzhanov, and A. I. Sattarov, Phys. Rev. {C 75}, 054003 (2007).
\bibitem{MES12} A. M. Mukhamedzhanov, V. Eremenko and A. I. Sattarov, Phys. Rev. C 86, 034001 (2012).
\bibitem{alt2002} E. O. Alt, A. M. Mukhamedzhanov, M. M. Nishonov, and A. I. Sattarov, Phys. Rev. { C 65}, 064613 (2002).
\bibitem{deltuva2009} A. Deltuva and A. C. Fonseca, Phys. Rev. { C 79}, 014606 (2009).
\bibitem{deltuva2009b} A. Deltuva, Phys. Rev. {C 79}, 054603 (2009).
\bibitem{cattapan1} G. Cattapan, G. Pisent  and V. Vanzani, Nucl. Phys. A 241, 204 (1975).
\bibitem{cattapan2} G. Cattapan, E. Mglione, G. Pisent and V. Vanzani, Nucl. Phys. A 296, 263 (1978).
\bibitem{miyagawa} K. Miyagawa  and Y. Koike, Prog. Theor. Phys. 82, 329 (1989).
\bibitem{alt_02} E.O. Alt, A.M. Mukhamedzhanov, M.M. Nishonov and A.I. Sattarov, Phys. Rev. C 65, 064613 (2002).
\bibitem{a1} B. Hoop, Jr., and H. H. Barschall, Nucl. Phys. 83, 65 (1966).
\bibitem{Shub16} Shubhchintak, C.A. Bertulani, A.M. Mukhamedzhanov and A. Kruppa, arXiv:1605.03206 (2016).
\bibitem{a2} G. L. Morgan and R. L. Walter, Phys. Rev. { 168}, 1114 (1968).
\bibitem{a3} A. C. L. Barnard, C. M. Jones and J. L. Weil, Nucl. Phys. { 50}, 604 (1964).
\bibitem{a4} L. Brown, W. Haeberli and W. Trilchslin, Nucl. Phys. A {90}, 339 (1967).
\bibitem{a5} G. Pisent and A. M. Saruis, Nuovo Cim. {28}, 600 (1963).
\bibitem{a6} W. Tr{\"a}chslin and L. Brown, Nucl. Phys. A {101}, 273 (1967).
\bibitem{a7} J. L. Fowler and H. 0. Cohn, Phys. Rev. {109}, 89 (1958).
\bibitem{a8} C. H. Johnson and J. L. Fowler, Phys. Rev. {162}, 890 (1967).
\bibitem{a9} S. R. Salisbury and H. T. Richards, Phys. Rev. {126}, 2147 (1962)
\bibitem{a10} R. A. Blue and W. Haeberli, Phys. Rev. {137}, 284 (1965).


\end{thebibliography}

\end{document}